%% file: arxiv_main.tex
\documentclass[lettersize,journal]{IEEEtran}

\def\BibTeX{{\rm B\kern-.05em{\sc i\kern-.025em b}\kern-.08em
    T\kern-.1667em\lower.7ex\hbox{E}\kern-.125emX}}

\usepackage{rotating}
\usepackage{amssymb}
\usepackage{algorithm}
\usepackage{algpseudocode}
\usepackage{color}
\usepackage{epsf}
\usepackage{psfrag}
\usepackage{epsfig}
\usepackage{multicol}
\usepackage{multirow}
\usepackage{makecell}
\usepackage{amsfonts}
\usepackage{textcomp}
\usepackage{comment}
\usepackage{footnote}
\usepackage{tablefootnote}
\usepackage[flushleft]{threeparttable}
\usepackage{paralist}
\usepackage{mdwlist}
\usepackage{amssymb}
\usepackage{amsmath}
\usepackage{url}
\usepackage{verbatim}
\usepackage{alltt}
\usepackage{balance}
\usepackage{multirow}
\usepackage{epstopdf}
\usepackage{lipsum}
\usepackage{float}
\usepackage{booktabs}
\usepackage{slashbox}
\usepackage[table]{xcolor}
\usepackage{array}
\usepackage{boldline}
\usepackage{caption,setspace}
\usepackage{mathtools}
\usepackage[table]{xcolor}
\usepackage{csquotes}

\usepackage{dsfont}
\usepackage{xcolor}

\usepackage{amsthm}

\usepackage{stackengine}[2013-10-15]

\usepackage[T1]{fontenc}
\usepackage{frcursive}
\usepackage{calligra}
\usepackage{wedn}
\usepackage{bm}
\usepackage{aurical}
\usepackage{upgreek}

\definecolor{darkgreen}{RGB}{0,204,0}

\usepackage{cite}
\usepackage{adjustbox}

\usepackage{esint}
\usepackage{tipa}
\usepackage{soul}
\usepackage{tikz}
\usepackage{listings}
\usepackage{multicol}
\usepackage{longtable} 
\usepackage{enumitem}
\usepackage{mdframed}

\newcolumntype{?}{!{\vrule width 1pt}}
\newcolumntype{+}{!{\vrule width 1.25pt}}

\def\hlineb#1{%
\noalign{\ifnum0=`}\fi\hrule \@height #1 %
\futurelet\reserved@a\@xhline}

\usepackage{cancel}
\usepackage{subcaption}

\usepackage{pgfplots}
\pgfplotsset{compat=1.18} 
\usepackage{mathrsfs}
\usepackage{pifont}

\usepackage[most]{tcolorbox}
\tcbuselibrary{listingsutf8}

\newcolumntype{L}[1]{>{\raggedright\arraybackslash}m{#1}}
\newcolumntype{C}[1]{>{\centering\arraybackslash}m{#1}}

\begin{document}

\title{Adaptive Modality Reliability Diagnosis and Restoration for Robust Multimodal Intent Recognition}

\author{
        Suraj Kumar, ~\IEEEmembership{Graduate Student Member, IEEE},
        Mohnish Raj,
        Soumi Chattopadhayay, ~\IEEEmembership{Senior Member, IEEE},
        Chandranath Adak, ~\IEEEmembership{Senior Member, IEEE},
        Ayan Dutta
\thanks{
  Corresponding author: Soumi Chattopadhayay, email: soumi@iiti.ac.in.
  This work has been submitted to a prominent venue for possible publication. Copyright may be transferred without notice, after which this version may no longer be accessible.
}
}

\maketitle

\begin{abstract}

    Multimodal intent recognition combines linguistic, acoustic, and visual evidence, but individual modalities may be noisy, missing, semantically conflicting, or disproportionately dominant. Existing methods typically infer modality importance implicitly and either reweight or suppress unreliable inputs, without determining whether a degraded modality can be repaired and subsequently trusted. We propose PRIME (Precision-weighted Reliability Inference and Modality rEstoration), a closed-loop reliability-guided framework that jointly diagnoses, restores, and reassesses modality quality at the sample level. PRIME represents the weakness of each modality through a contextual log-variance estimated from complementary diagnostic evidence, including predictive confidence,
    epistemic disagreement, cross-modal consensus, and feature degeneracy. Because modality-reliability annotations are unavailable, the estimator is explicitly trained using controlled modality corruption with known degradation severity, together with a heteroscedastic uncertainty objective. Rather than directly discarding an unreliable modality, PRIME uses its estimated weakness to control a prototype-conditioned variational restoration module that reconstructs the degraded representation from complementary modalities. Crucially, reliability is re-estimated after restoration, allowing the model to determine whether the repaired representation has become sufficiently trustworthy to contribute to prediction. The resulting post-restoration precisions are used for inverse-variance multimodal fusion. Experiments on multimodal intent-recognition benchmarks show that PRIME maintains competitive clean-data performance while improving robustness under missing, noisy, conflicting, and modality-imbalanced conditions.

\end{abstract}

\input{1_intro}
\input{2_related_work_short}

\input{3_method_new}

\input{4_experiments}
\input{5_conclusion}

\bibliographystyle{ieeetr}
\bibliography{aaai2027}

\section*{Supplementary Appendix}
Appendices are provided at \url{https://github.com/csksuraj17/PRIME}

\end{document}

%% file: 1_intro.tex
\section{Introduction} 
\label{sec:intro}
Multimodal Intent Recognition (MIR) is an emerging computing task that recognizes communicative intent from language, speech, and visual cues, with applications in real-time conversation analysis, emotion understanding, and healthcare \cite{zhu2025survey}. Early unimodal approaches (e.g., text-only) are insufficient for capturing rich human interactions \cite{zhao2025deep_survey}. While recent multimodal methods leverage textual, acoustic, and visual cues to improve intent recognition~\cite{TCLMAP_aaai_2024,MVCL_DAF_aaai2025}, modeling heterogeneous modalities in real-world settings remains challenging, as missing, noisy, uninformative, or conflicting modalities can induce negative cross-modal transfer and degrade overall performance rather than contribute complementary information \cite{missing_modality_SMIL_AAAI_21,yang2025uncertain}.

Existing studies \cite{mullick2025emnlp,MVCL_DAF_aaai2025} have shown that the textual modality provides the strongest semantic evidence and consequently dominates acoustic and visual signals. In scenarios where certain modalities are unreliable or partially missing, suppressing or discarding them may result in information loss and reduced robustness. Instead, selectively compensating weaker modalities with complementary cues offers a more principled alternative. To enhance the semantic representations of non-textual modalities, \cite{LiPLC25} proposed a text-guided cross-modal attention mechanism with text-centric adaptive fusion gating. To improve robustness under uncertain or missing modalities, \cite{mine_yang2025uncertain_cvpr} exploited emotion--intention correlations through implicit label reasoning and modality-asynchronous prompting. To address modality imbalance caused by dominant modalities, \cite{ARL_NIPS_2025} proposed a dual-path recalibration framework with sample-wise modality calibration and encoder-level weight recalibration. Similarly, \cite{cdpr_icmr_2026} addressed cross-modal semantic inconsistencies through a cognitive dual-pathway reasoning framework that jointly models consensus and conflict for adaptive trust recalibration and modality fusion. Despite these advances, modality reliability is typically learned implicitly, as existing MIR benchmarks do not provide modality-level reliability annotations to explicitly supervise reliability estimation. Consequently, current methods rely on heuristic objectives, which may produce unreliable reliability estimates under complex real-world conditions.

To address these limitations, we revisit a fundamental question: what constitutes a reliable modality? We argue that reliable modality estimation should satisfy three key properties. First, confidence alone is insufficient, as a modality can be confidently incorrect; reliability therefore requires calibrated evidence. Second, modality reliability is inherently context-dependent, since whether a modality is informative depends on its agreement with the remaining modalities rather than on its individual confidence. Third, reliability should be explicitly learnable, rather than emerging implicitly from the intent classification objective, which often biases learning toward text-dominant solutions on existing MIR benchmarks. 

Motivated by these insights, we propose PRIME (Precision-weighted Reliability Inference and Modality rEstoration), a sample-specific framework that explicitly estimates and exploits modality reliability. PRIME first characterizes each modality using complementary diagnostic cues that capture confidence, uncertainty, 
and cross-modal agreement, and jointly reasons over them through a contextual reliability router to produce reliability-aware modality representations. Rather than suppressing unreliable modalities, PRIME restores them using complementary information from the remaining modalities, preserving useful semantic cues while mitigating modality degradation. Since explicit reliability annotations are unavailable, we further introduce a self-supervised reliability learning objective that generates supervisory signals through controlled modality corruption, enabling accurate reliability estimation and robust modality restoration.

\noindent\textbf{To summarize, our contributions are as follows:}
\begin{itemize}
     \item We propose \textbf{PRIME}, a sample-specific, reliability-aware framework for multimodal intent recognition that explicitly estimates modality reliability and adaptively restores unreliable modalities in the presence of modality imbalance, noise, and missing signals, without relying on expensive cross-modal fusion.
    
    \item We introduce a contextual reliability estimation mechanism that jointly reasons over complementary diagnostic cues to infer modality reliability. To overcome the lack of reliability annotations, we formulate reliability learning as a self-supervised corruption prediction task, further regularized by a heteroscedastic uncertainty objective.
    
    \item We propose a prototype-conditioned modality restoration module that reconstructs degraded modality representations from complementary modalities and integrates them through reliability-guided inverse-variance fusion, preserving informative cues instead of suppressing weak modalities.
    
    \item Extensive experiments on standard multimodal intent recognition benchmarks demonstrate that PRIME consistently outperforms state-of-the-art methods and exhibits superior robustness under noisy, missing, conflicting, and text-dominant multimodal scenarios.
\end{itemize}

%% file: 2_related_work_short.tex
\section{Related Work}
\label{sec:related_work}

\subsection{Multimodal Intent Recognition}

Multimodal intent recognition (MIR) has evolved from early text- and vision-based models toward frameworks that jointly exploit textual, acoustic, and visual information. Initial multimodal approaches integrated textual and acoustic cues through shared representation learning \cite{huang2022mtl,seo2022integration}, while prototype-based methods further improved cross-modal semantic alignment \cite{MuProCL}. Earlier unimodal studies also explored prototype learning and weakly supervised vision-language modeling for intent recognition \cite{PIP-NET_TMM25,IntCLIP_eccv2024}. More recently, multimodal large language models (MLLMs) have substantially advanced semantic reasoning in MIR. Representative methods include LGSRR \cite{LGSRR_emnlp_2025}, which leverages LLM-guided reasoning, HIER \cite{hier_cvpr_26}, which performs hierarchical concept reasoning, and MIntOOD \cite{MIntOOD}, which jointly addresses intent recognition and out-of-distribution detection. Although these approaches significantly improve multimodal understanding, they generally assume that all modalities provide equally reliable evidence for each sample.

\subsection{Fusion and Modality Reliability in MIR}

Recent studies have shown that MIR is highly susceptible to modality bias, where dominant modalities, particularly text, can overshadow complementary visual and acoustic information \cite{mullick2025emnlp}. To address this issue, numerous multimodal fusion strategies have been proposed, including attention-based fusion \cite{huang2023effective,liu2021hierarchical}, frequency-aware fusion \cite{gong2025wdmir}, modality-aware prompting \cite{zhou2024token}, contrastive fusion \cite{TCLMAP_aaai_2024}, dynamic attention mechanisms \cite{MVCL_DAF_aaai2025}, and adaptive modality recalibration \cite{ARL_NIPS_2025}. While these methods improve cross-modal interaction, modality importance is typically learned implicitly from the downstream task, rather than through an explicit estimate of modality reliability.

A related line of research focuses on improving robustness under noisy, missing, or conflicting modalities. Information bottleneck methods, such as InMu-Net \cite{InMuNet_MM_2024} and SeD-UD \cite{SedUD_2026_CVPR}, suppress redundant or noisy information through adaptive feature compression. Other approaches explicitly estimate modality confidence using feature statistics or prediction consistency, including ECFMIR \cite{wcfmir_aaai26} and CDPR \cite{cdpr_icmr_2026}, while CR-3WD \cite{cr3wd_esa_26} regenerates missing modality representations before prediction. Despite these advances, existing methods either suppress unreliable modalities or derive reliability indirectly from feature statistics or classification objectives, providing no explicit supervision for modality reliability and limited capability to recover degraded modality representations.

\input{figures/main_arch}

\subsection{Positioning of Our Work}
PRIME explicitly models modality reliability as a sample-specific latent variable through self-supervised corruption, enabling reliability to be learned directly rather than inferred from classification confidence or feature statistics. Reliability is estimated from complementary diagnostic cues, including energy-based confidence, 
epistemic disagreement, divergence from multimodal consensus, and feature degeneracy, which are aggregated by a permutation-equivariant contextual router to guide prototype-conditioned restoration before fusion. Unlike existing methods that primarily reweight or suppress unreliable modalities, PRIME repairs weak modality representations, improving robustness under modality imbalance, missing modalities, noise, and cross-modal conflicts.

%% file: figures/main_arch.tex
\begin{figure*}[htbp]
    \centering
    \includegraphics[width=18cm, height=6.5cm]{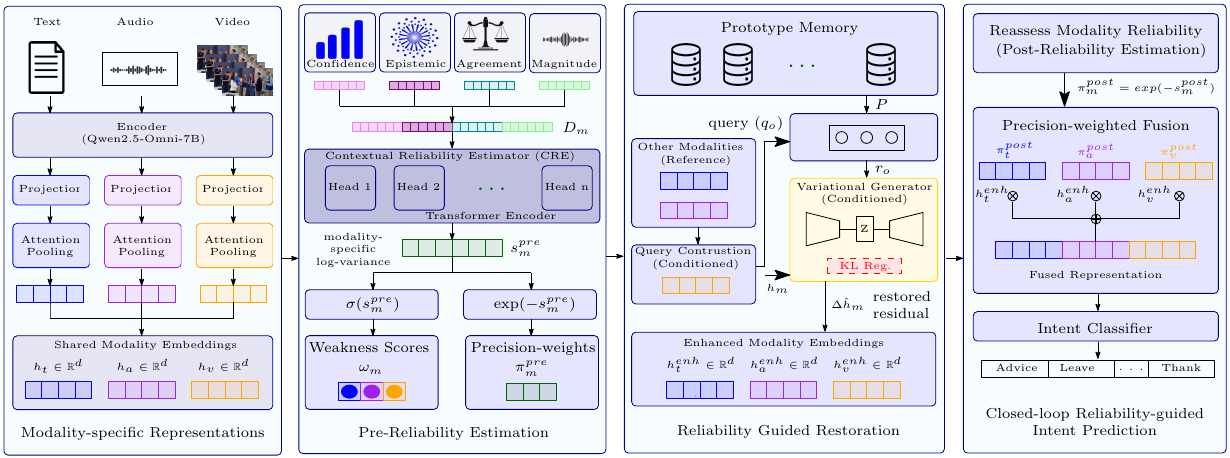}
    \caption{\textbf{Overview of PRIME.} Given multimodal embeddings, PRIME first estimates modality reliability using complementary diagnostic probes and a contextual reliability estimator. Reliability then guides prototype-assisted residual restoration, after which enhanced representations are re-evaluated to obtain updated precisions for inverse-variance fusion and intent prediction.}
    \label{fig:arch}
\end{figure*}

%% file: 3_method_new.tex
\section{Methodology}

Figure~\ref{fig:arch} illustrates the overall architecture of PRIME. Existing multimodal intent recognition methods treat modality reliability as a by-product of feature learning. PRIME instead models reliability as an explicit latent variable that evolves throughout the inference process. Reliability is first estimated from complementary diagnostic evidence, then used to guide cross-modal restoration, and finally re-estimated on the enhanced representations before reliability-aware fusion.

\subsection{Problem Formulation}
We consider multimodal intent recognition (MIR), where each utterance consists of text $T$, audio $A$, and video $V$ associated with an intent label $c\in C$. Given a training set \(
\mathcal{D}
=
\{(T_i,A_i,V_i,c_i)\}_{i=1}^{N},
\)
the objective is to learn a mapping
\(
f:(T,A,V)\rightarrow c.
\)

Unlike conventional multimodal fusion methods, PRIME explicitly models the reliability of each modality before prediction. Since modality reliability is not annotated in existing MIR datasets, we generate supervision through controlled modality corruption during training. This enables the model to jointly learn reliability estimation and cross-modal restoration without requiring additional annotations.

For each modality $m\in\{t,a,v\}$, the reliability estimator predicts a log-variance \(s_m\in\mathbb{R},\) which parameterizes both the modality weakness
\(\omega_m=\sigma(s_m),\) and its corresponding precision \(\pi_m=\exp(-s_m).\)

The weakness score supervises restoration, while the precision determines both source selection during restoration and reliability-aware multimodal fusion.


\subsection{Reliability Estimation}

PRIME begins by estimating the reliability of each modality from complementary diagnostic cues. Rather than relying solely on prediction confidence, the proposed estimator jointly considers confidence, 
epistemic uncertainty, cross-modal agreement, and representation quality to produce a sample-specific reliability estimate.


\subsubsection{Initial Feature Extraction}
Given an input utterance, modality-specific representations are extracted using the frozen Qwen2.5-Omni-7B encoder~\cite{qwen25omnitechnicalreport}. For each modality \(m\in\{t,a,v\},\) the encoder produces a sequence 
\(
    X_m=\{x_{m,i}\}_{i=1}^{l_m},
    ~~
    x_{m,i}\in\mathbb{R}^{d_m},
\)
which is projected into a shared embedding space,
\begin{equation} \small
    H_m
    =
    \mathrm{Proj}_m(X_m)
    \in
    \mathbb{R}^{l_m\times d}.
\end{equation}
To obtain a fixed-length representation while ignoring padded positions, masked attentive pooling is applied,
\begin{equation} \small
    \small
    h_m
    =
    \sum_i
    a_m^{(i)}
    H_m^{(i)}
    \in
    \mathbb{R}^{d}, ~~
    a_m^{(i)}
    =
    \frac{
    \exp(w^\top H_m^{(i)})
    \mathbf{1}_m^{(i)}
    }{
    \sum_j
    \exp(w^\top H_m^{(j)})
    \mathbf{1}_m^{(j)}
    },
\end{equation}
%
%
where $\mathbf{1}_m$ denotes the validity mask. Each pooled representation is further passed through a lightweight auxiliary classifier,
\(
z_m=W_mh_m,
~~
p_m=\mathrm{softmax}(z_m),
\)
whose predictions provide complementary evidence for reliability estimation.

\subsubsection{Diagnostic Reliability Probes}

A reliable modality should produce confident predictions, remain close to the underlying class manifold, agree with complementary modalities, and maintain stable feature representations. To capture these properties, PRIME computes 
four complementary diagnostic measures for every modality.

\paragraph{Prediction confidence.}
Prediction confidence is measured using the energy score
\begin{equation} \small
    F_{\mathrm{en}}(z_m)
    =
    -
    \log
    \sum_c
    \exp
    \left(
    z_m^{(c)}
    \right).
\end{equation}
Unlike maximum softmax probability, energy provides a more reliable confidence estimate under uncertain predictions.


\paragraph{Epistemic uncertainty.}
Following lightweight ensemble estimation, \(K\) linear prediction heads are attached to each modality representation, and the unbiased variance of their predicted probabilities is computed as
%
\begin{equation} \small
    F_{\mathrm{ep}}(h_m)
    =
    \sum_c
    \mathrm{Var}_k
    \left[
    \mathrm{softmax}
    (g_m^k(h_m))^{(c)}
    \right].
\end{equation}
Higher disagreement indicates greater epistemic uncertainty.

\paragraph{Cross-modal agreement.}
Because modality quality is relative to the remaining modalities, we compare each prediction with the multimodal consensus,
\(
    \bar p
    =
    \frac{1}{3}
    \sum_m
    p_m,
\)
using Jensen-Shannon divergence,
\begin{equation} \small
    F_{\mathrm{co}}
    (p_m,\bar p)
    =
    \mathrm{JS}(p_m||\bar p).
\end{equation}

Large divergence suggests that a modality provides evidence inconsistent with the remaining modalities.

\paragraph{Representation quality.}
Finally, feature degeneracy is estimated through the embedding norm,
\(
    F_{\mathrm{de}}
    (h_m)
    = ||h_m||_2,
\)
which provides a lightweight indicator of collapsed or near-zero representations.

The four
diagnostics are concatenated and normalized to obtain the reliability descriptor
\begin{equation} \small
    D_m
    =
    \mathrm{BN}
    \Big(
    F_{\mathrm{en}}
    ||
    F_{\mathrm{ep}}
    ||
    F_{\mathrm{co}}
    ||
    F_{\mathrm{de}}
    \Big)
    \in
    \mathbb{R}^{4},
\end{equation}
which summarizes the reliability evidence for modality $m$ and serves as input to the contextual reliability estimator.


\subsubsection{Contextual Reliability Estimator}

The diagnostic probes characterize each modality independently, whereas modality reliability is inherently relative to the remaining modalities within the same sample. PRIME therefore estimates reliability jointly across modalities using a lightweight Transformer-based Contextual Reliability Estimator (CRE).

The diagnostic descriptors are first stacked into
\(
    D=\{D_m\}_{m\in\{t,a,v\}}
    \in
    \mathbb{R}^{M\times4},
\)
embedded through a linear projection with learnable modality embeddings, and processed by a Transformer encoder operating along the modality dimension. A prediction head estimates a modality-specific log-variance,
\begin{equation} \small
    \{s_m\}_m
    =
    \mathrm{Head}
    \left(
    \mathrm{Enc}
    (\{D_m\}_m)
    \right),
    \qquad
    s_m\in\mathbb{R}.
\end{equation}
Self-attention enables each modality to be evaluated in the context of the remaining modalities instead of using independent reliability estimates. The predicted log-variance is converted into the modality weakness
\(
\omega_m=\sigma(s_m),
\)
and the corresponding precision
\(
\pi_m^{pre}=\exp(-s_m),
\)
which are subsequently used to guide restoration and multimodal fusion.


\subsection{Reliability-guided Restoration}
The estimated reliability determines not only how modalities are fused but also how they are restored. Rather than suppressing unreliable modalities before fusion, PRIME reconstructs degraded representations using complementary modalities that are themselves estimated to be reliable. This enables the model to recover missing intent-relevant information while limiting error propagation from unreliable sources.


\subsubsection{Prototype-guided Cross-modal Restoration}
For each target modality $m$, restoration is generated independently from every complementary source modality $o\neq m$. A semantic query
\(
q_o=Q(h_o)
\)
is first computed from the source representation and used to retrieve intent-aware semantic information from a learnable prototype memory
\(
P
\in
\mathbb{R}^{N_p\times d},
\)
where $N_p=\kappa \times |C|$ denotes $\kappa$ prototypes for each intent classes. Prototype retrieval is performed through attention,
\begin{equation} \small
    {a}_o
    =
    \mathrm{softmax}
    \left(
    \frac{q_oP^\top}{\tau}
    \right),
    \qquad
    r_o
    =
    {a}_oP,
\end{equation}
where $\tau$ is the temperature parameter.
Instead of reconstructing an entire modality, PRIME predicts only the residual required to recover the missing information. A latent variable is inferred from the source representation,
\begin{align}
    \mu_o,&~\log\sigma_o^2
    =
    \Phi(h_o),\\
    z_o
    &=
    \begin{cases}
    \mu_o+\sigma_o\odot\xi,
    &
    \text{training},
    \\
    \mu_o,
    &
    \text{inference},
    \end{cases}
\end{align}
\(\xi \sim \mathcal{N}(0, I)\), and combined with the retrieved prototype to predict the source-specific residual,
\begin{equation} \small
    \delta_o
    =
    \mathrm{Dec}
    \left(
    [r_o;
    h_m;
    z_o]
    \right).
\end{equation}
Learning the residual preserves reliable information already present in the target modality while allowing the generator to focus on recovering only the degraded components. The variational latent captures the one-to-many nature of cross-modal generation and is regularized through
\begin{equation} \small
    \mathcal{L}_{\mathrm{kl}}
    =
    -\frac12
    \sum_j
    \left(
    1+
    \log\sigma_{o,j}^2
    -
    \mu_{o,j}^2
    -
    \sigma_{o,j}^2
    \right).
\end{equation}


\subsubsection{Reliability-aware Source Selection}
Complementary modalities are beneficial only when they provide reliable guidance. To prevent unreliable modalities from introducing additional noise, PRIME selects restoration sources according to the estimated precision.

For each source modality,
\begin{equation} \small
    \tau_o
    =
    \mathrm{median}_b
    \left(
    \pi_o^{(b)}
    \right),
    \qquad
    \gamma_o^{(b)}
    =
    \mathbb{1}
    \left[
    \pi_o^{(b)}
    \ge
    \tau_o
    \right],
\end{equation}

where the batch median provides an adaptive reliability threshold. Only the selected sources contribute to the restored residual,
\begin{equation} \small
    \Delta\hat h_m
    =
    \frac{
    \sum_{o\neq m}
    \gamma_o
    \odot
    \delta_o
    }{
    \sum_{o\neq m}
    \gamma_o+\epsilon
    },
\end{equation}
where $\epsilon>0$ is used for numerical stability. \(\Delta\hat h_m\) is then injected into the target representation,
\begin{equation} \small
    h_m^{\mathrm{enh}}
    =
    h_m
    +
    \Delta\hat h_m.
\end{equation}
This residual formulation naturally preserves reliable representations, since clean modalities produce near-zero corrections, while degraded modalities receive larger updates.


\subsubsection{Closed-loop Reliability Update}
Restoration changes the quality of modality representations and therefore alters their reliability. PRIME consequently re-evaluates every enhanced representation using the same diagnostic probes and contextual reliability estimator. The updated reliability estimates determine whether restoration has sufficiently improved a modality before it contributes to multimodal fusion, establishing a closed-loop interaction between reliability estimation and restoration.

\subsection{Reliability-aware Fusion}
Restoration modifies the quality of modality representations, making the initial reliability estimates no longer representative. PRIME therefore performs a second reliability estimation on the enhanced representations using the diagnostic probes and contextual reliability estimator. The updated log-variances $s_m^{\mathrm{post}}$ are converted into precisions,
\begin{align}
    \pi^{post}_m
    &=
    \exp\left(-s_m^{\mathrm{post}}\right), \qquad
    \beta_m
    =
    \frac{\pi^{post}_m}
    {\sum_{m'}\pi_{m'}^{post}},
\end{align}

which determines the contribution of each modality during fusion. The final multimodal representation is obtained as
\begin{equation} \small
    H
    =
    \sum_{m\in\{t,a,v\}}
    \beta_m
    h_m^{\mathrm{enh}},
\end{equation}
where more reliable modalities receive larger fusion weights. The fused representation $H$ is then passed to a lightweight multilayer classifier for intent prediction.


\subsection{Optimization}

PRIME is optimized end-to-end using objectives for intent prediction, reliability estimation, and cross-modal restoration. \textbf{Intent prediction} is supervised by the standard class-balanced cross-entropy loss $\mathcal{L}_{\mathrm{cls}}$. 
To learn the \textbf{Reliability}, PRIME combines heteroscedastic uncertainty learning with self-supervised corruption severity estimation. Let $i$ index samples within a mini-batch $\mathcal{B}$ and $m$ index modalities. Following learned loss attenuation, the auxiliary classifier of each modality is optimized using
\begin{equation} \small
    \mathcal{L}_{\mathrm{het}}
    =
    \frac{1}{|\mathcal{B}|}\sum_{i}
    \frac13
    \sum_m
    \left[
    e^{-s_{m,i}} \mathrm{CE}(z_{m,i},c_i)
    +
    \frac12
    s_{m,i}
    \right].
\end{equation}
%
Since modality reliability is unavailable in existing datasets, we synthesize supervision by randomly corrupting each modality during training with sampled severity 
$\mathcal{M}_m$
The estimated weakness is then trained to regress the injected corruption level,
\begin{equation} \small
    \mathcal{L}_{\mathrm{sev}}
    =
    \frac{1}{|\mathcal{B}|}\sum_{i}
    \frac1{|\mathcal{E}|}
    \sum_{m\in\mathcal{E}}
    (\omega_{m,i}-\mathcal{M}_{m,i})^2;
    \quad\mathcal{E}=\{t,a,v\}.
\end{equation}
%
Further, the \textbf{Restoration} is supervised through residual reconstruction, semantic consistency, prototype regularization, and variational regularization. The synthetic corruption process simultaneously provides paired clean representations, enabling direct supervision of the restoration module. The predicted residual is optimized using
\begin{equation} \small
    \mathcal{L}_{\mathrm{res}}
    =
    \frac{1}{|\mathcal{B}|}\sum_{i}
    \frac1{|\mathcal{E}|}
    \sum_{m\in\mathcal{E}}
    ||
    \Delta\hat h_{m,i}
    -
    \mathrm{sg}
    \left[
    h_{m,i}^{\mathrm{clean}}
    -
    h_{m,i}
    \right]
    ||_2^2 .
\end{equation}
%
where $\mathrm{sg}[\cdot]$ denotes the stop-gradient operator. To preserve semantic discriminability after restoration, the enhanced representations are additionally constrained by
\begin{equation} \small
    \mathcal{L}_{\mathrm{gc}}
    =
    \frac{1}{|\mathcal{B}|}\sum_{i}
    \frac1{|\mathcal{E}|}
    \sum_{m\in\mathcal{E}}
    \mathrm{CE}
    \left(
    \mathrm{aux}_m
    (h_{m,i}+\Delta\hat h_{m,i}),
    c_i
    \right).
\end{equation}
%
Finally, the prototype memory is regularized using a load-balancing objective \(\mathcal{L}_{\mathrm{lb}}\) together with a class-consistency constraint \(\mathcal{L}_{\mathrm{pc}}\) that maximizes the entropy of mean prototype usage $\bar a_j = \frac{1}{|\mathcal{B}|}\sum_i a_{j,i}$ for $j$-th class
\begin{equation} 
    \small
    \mathcal{L}_{\mathrm{lb}}
    =
    \log N_p
    +
    \sum_j
    \bar a_j
    \log \bar a_j, \quad
    \mathcal{L}_{\mathrm{pc}}
    =
    -
    \log
    \left(
    \sum_{j:y_j=c}
    \bar a_j
    \right).
\end{equation}
The complete optimization objective is
\begin{equation} \small
    \begin{aligned}
        \mathcal{L}
        =
        &
        \lambda_{\mathrm{cls}} \mathcal{L}_{\mathrm{cls}}
        +
        \lambda_{\mathrm{het}}
        \mathcal{L}_{\mathrm{het}}
        +
        \lambda_{\mathrm{sev}}
        \mathcal{L}_{\mathrm{sev}}
        +
        \lambda_{\mathrm{res}}
        \mathcal{L}_{\mathrm{res}}
        +
        \lambda_{\mathrm{gc}}
        \mathcal{L}_{\mathrm{gc}}
        \\
        &
        +
        \lambda_{\mathrm{lb}}
        \mathcal{L}_{\mathrm{lb}}
        +
        \lambda_{\mathrm{pc}}
        \mathcal{L}_{\mathrm{pc}}
        +
        \lambda_{\mathrm{kl}}
        \mathcal{L}_{\mathrm{kl}}.
    \end{aligned}
\end{equation}

%% file: 4_experiments.tex
\input{tables/dataset} 

\section{Experiments}\label{sec:experiments}

\input{tables/sota_table}

\subsection{Dataset Employed and Evaluation Metrics} We conduct experiments on three multimodal conversational benchmarks: single-turn-focused MIntRec~\cite{mintrec} 
and the multi-party in-scope subset of MIntRec2.0~\cite{mintrec2}.
All dataset statistics are summarized in Table~\ref{tab:dataset}. Following previous studies, we evaluate our framework using standard classification metrics: Accuracy (Acc), Precision (P), Recall (R), F1-score (F1), weighted precision (WP), and weighted F1-score (WF1).

\subsection{Implementation Details} 
PRIME is built upon the frozen Qwen2.5-Omni-7B backbone~\cite{qwen25omnitechnicalreport}, which is adapted via LoRA before feature extraction and produces unified multimodal embeddings that are projected into a shared 256-dimensional space. The model is trained using AdamW with early stopping (patience = 5) based on the validation weighted F1-score. During training, a randomly selected modality is synthetically corrupted to supervise reliability estimation and cross-modal restoration. Results are reported on test data over five seeds (1-5). Additional details are given in the supplementary material.



\subsection{Main Results and Comparative Study}

We compare PRIME with eleven representative multimodal intent recognition methods, including classical fusion models (MISA~\cite{misa_MM_20}, MulT~\cite{mult_2019_acl}, MAG-BERT~\cite{kumar2024introducing}, TCL-MAP~\cite{TCLMAP_aaai_2024}, MVCL-DAF~\cite{MVCL_DAF_aaai2025}, and WDMIR~\cite{gong2025wdmir}), reliability-aware approaches (MIntOOD~\cite{MIntOOD}, ECFMIR~\cite{wcfmir_aaai26}, CDPR~\cite{cdpr_icmr_2026}, and CR-3WD~\cite{cr3wd_esa_26}), and the recent MLLM-based framework HIER~\cite{hier_cvpr_26}.

Table~\ref{tab:main_results} compares PRIME with representative multimodal intent recognition methods. PRIME consistently achieves the best performance on both benchmarks, surpassing the previous state-of-the-art method, HIER, while maintaining clear margins over conventional fusion-based approaches.

On MIntRec, PRIME achieves the highest performance across all reported metrics, with particularly notable improvements in recall, macro-F1, and weighted F1, indicating more robust recognition across diverse intent categories. The performance margin is even more pronounced over conventional fusion-based methods, highlighting the advantage of explicitly modeling modality reliability rather than relying solely on feature interaction. On the more challenging MIntRec2.0 benchmark, PRIME remains consistent and attains \(1.22\%\) improvement on WF1, suggesting that the proposed reliability-aware framework generalizes well to larger intent taxonomies and more diverse conversational settings.

These improvements are consistent with the design of PRIME. Existing methods either estimate modality importance implicitly through feature fusion or perform a single reliability assessment before prediction. In contrast, PRIME explicitly learns modality reliability through self-supervised corruption, exploits the estimated reliability to guide cross-modal restoration, and re-estimates reliability before precision-weighted fusion. This closed-loop \emph{diagnose-restore-reassess} strategy enables degraded modalities to be repaired before participating in prediction, leading to more robust multimodal representations and consistently stronger intent recognition performance.

\subsection{Ablation Study}
To assess the impact of each component in PRIME, we conduct ablation studies on the modules, diagnostic probes, modality restoration, and training objectives. 

\textbf{Impact of core components in PRIME.}
We evaluate the contribution of each core component in PRIME, as shown in Table~\ref{tab:module}. Compared with the encoder-only baseline, the complete framework achieves substantial improvements across all evaluation metrics, showing that the performance gains arise from the proposed reliability learning framework rather than the backbone alone. Removing any individual component consistently degrades performance, confirming that each contributes to the overall design.

Removing either the contextual reliability estimator or the generative restoration module degrades performance by an average of $1.69\%$ and $1.64\%$, respectively, across the five evaluation metrics, highlighting the importance of contextual reliability estimation and reliability-guided modality restoration. Replacing precision-weighted fusion with uniform averaging reduces performance by an average of $1.07\%$, indicating that reliability-aware fusion provides additional benefits once enhanced representations are available. Overall, these results demonstrate that reliability estimation, restoration, and precision-weighted fusion contribute complementary improvements, validating the proposed closed-loop \emph{diagnose-restore-reassess} framework.


\input{tables/module_study}

\textbf{Impact of Diagnostic Probes.} 
We next analyze the contribution of the proposed diagnostic probes from two complementary perspectives; the results are reported in Table~\ref{tab:diag}. The leave-one-out study shows that removing any individual probe consistently degrades performance, confirming that each contributes useful information for modality reliability estimation. Removing the energy-based confidence or epistemic uncertainty probe results in slightly larger degradations than the remaining probes, highlighting their complementary contribution to reliability estimation.

The single-probe study further shows that no individual diagnostic matches the performance of the complete framework. Confidence achieves the highest standalone performance, whereas epistemic uncertainty and cross-modal consensus are comparatively less effective in isolation. However, combining all diagnostic cues consistently outperforms any individual diagnostic, demonstrating that reliable modality estimation benefits from integrating complementary sources of evidence rather than relying on a single confidence- or feature-based heuristic.
\input{tables/diagnostics}

\textbf{Benefits of Modality Restoration.}
Table~\ref{tab:modality} investigates which modalities benefit from generative restoration. Restoring all three modalities consistently achieves the best performance, whereas restoring only a single modality or excluding either text or video results in lower performance. This indicates that reliability-guided restoration provides complementary benefits across modalities.

An interesting trend emerges from the leave-one-out study. Excluding either text or video restoration leads to comparable performance degradation, whereas removing audio restoration has a smaller effect. This suggests that text and video restoration contribute more to the final prediction, while audio provides complementary information. Together, these results support applying reliability-guided restoration to all modalities rather than only a subset.



\input{tables/modality_study}
\input{tables/loss_study}
\textbf{Study of learning objectives.}
We next examine the contribution of each training objective, as shown in Table~\ref{tab:loss}. Removing any loss consistently degrades performance, indicating that the proposed objectives provide complementary supervision throughout the learning process. Omitting either the heteroscedastic loss ($\mathcal{L}_{\text{het}}$) or the corruption-severity supervision ($\mathcal{L}_{\text{sev}}$) results in comparatively larger performance degradations, highlighting the importance of explicitly learning modality reliability. The restoration objectives, including residual reconstruction ($\mathcal{L}_{\text{res}}$), semantic consistency ($\mathcal{L}_{\text{gc}}$), and variational regularization ($\mathcal{L}_{\text{kl}}$), also contribute consistently, indicating that reliable modality restoration benefits from dedicated supervision. In contrast, removing the prototype regularization losses ($\mathcal{L}_{\text{lb}}$ and $\mathcal{L}_{\text{pc}}$) leads to relatively smaller degradations, suggesting that they provide complementary regularization during training.

Training with only the standard classification loss also yields lower performance than the complete objective, confirming that PRIME benefits from jointly optimizing reliability estimation, modality restoration, and intent classification instead of relying on classification supervision alone.





Furthermore, the supplementary material includes hyperparameter sensitivity, class-wise performance, qualitative examples, failure case analysis, statistical significance testing, and runtime and computational complexity analyses.

%% file: tables/dataset.tex
\begin{table}[b]
    \centering
    \caption{Dataset Statistics.}
    \resizebox{0.48\textwidth}{!}{
        \begin{tabular}{lcccccc} \toprule
            \multirow{2}{*}{Dataset} & \multirow{2}{*}{\#C} & Avg. Utt. & Avg. Video &  \multirow{2}{*}{\#Train} & \multirow{2}{*}{\#Val} & \multirow{2}{*}{\#Test}  \\
    
            & & Len. & Len. (sec) & & & \\  \midrule
            MIntRec & 20 & 7.04 & 2.38 & 1334 & 445 & 445 \\
            MIntRec2.0 & 30 & 7.35 & 2.94 &  6165 & 1106 & 2003 \\ 
            \bottomrule
        \end{tabular}
    }
    \label{tab:dataset}
\end{table}

%% file: tables/sota_table.tex
\begin{table*}[t]
\centering
\caption{Performance comparison study. Best and second-best results are highlighted with \textbf{Bold} and \underline{underlined}, respectively.}
\label{tab:main_results}
\small
\resizebox{0.85\textwidth}{!}{
    \begin{tabular}{l|ccccc|ccccc}
    \toprule
    \multirow{2}{*}{\textbf{Methods}} &
    \multicolumn{5}{c|}{\textbf{MIntRec}} &
    \multicolumn{5}{c}{\textbf{MIntRec2.0}} \\
    \cmidrule(lr){2-6}
    \cmidrule(lr){7-11}
    &
    \textbf{Acc} & \textbf{R} & \textbf{F1} & \textbf{WP} & \textbf{WF1} &
    \textbf{Acc} & \textbf{R} & \textbf{F1} & \textbf{WP} & \textbf{WF1} \\ \midrule

    MISA & 72.29 &  73.48 & 69.24 & 70.85 & 69.32 & 55.16 & 49.92 & 49.51 & 57.06 & 55.05  \\
    
    MulT & 72.31 & 68.83 & 68.97 & 72.24 & 72.07  & 60.66 & 53.77 & 54.12 & 60.12 & 59.55  \\  
    
    MAG-BERT & 72.40 & 69.22 & 68.29 & 72.94 &  72.06  & 60.38 & 54.54 & 54.74 & 60.00 &  59.61  \\ 

    TCL-MAP & 73.17 & 69.99 & 68.92 & 72.97 &  72.66 & 58.24 & 52.41 & 52.25 & 57.55 &  57.24  \\  

    MIntOOD & 73.48	& 70.45	& 70.51	& 73.24	&  72.39 & 58.73& 51.20	& 52.40	& 58.34 & 58.03 \\ 
    
    MVCL-DAF & 73.63 & 70.11 & 70.41 & 74.31 & 73.57 & 59.64 &  53.24 & 53.41 & 58.57 & 58.67   \\ 

    
     ECFMIR &  73.26 &  67.08 &69.27 & 75.15 & 72.96 & 56.42 &  48.33 & 50.20 & 58.44 & 55.82  \\ 

    WDMIR & 75.06 & 72.65 & 72.76& 75.26 &  74.96 & 57.16 & 50.75 & 51.93 & 57.90 & 56.64 \\ 

    CDPR & 75.15 & 71.08 & 71.04 & 75.37 & 74.91 & 60.82 & 53.40 & 53.86 & 60.23 & 59.54  \\ 

    CR-3WD & 76.85 & 74.52 & 74.49 & 76.66 & 76.49 & 60.30 & 53.93 & 54.12 & 60.12 & 59.33 \\
    
    HIER & \underline{80.00} & \underline{77.11} & \underline{76.91} & \underline{80.67} & \underline{79.59} & \underline{64.15} & \underline{59.59} & \underline{60.31} & \underline{64.17} & \underline{63.79}  \\  
    
    \midrule
    \textbf{PRIME (ours)} & \textbf{81.57 } &  \textbf{79.85} & \textbf{79.18 }&\textbf{ 82.06 }& \textbf{81.53} & \textbf{64.66 } &\textbf{62.06} & \textbf{60.73} &\textbf{65.35} & \textbf{64.57}  \\
    &\scriptsize$\pm$0.50 & \scriptsize$\pm$0.84 &\scriptsize$\pm$0.71 & \scriptsize$\pm$0.58 & \scriptsize$\pm$0.59  & \scriptsize$\pm$0.92 & \scriptsize$\pm$1.24 & \scriptsize$\pm$0.82 & \scriptsize$\pm$0.82 & \scriptsize$\pm$0.60 \\
    \bottomrule
    \end{tabular}
    }
\end{table*}

%% file: tables/module_study.tex
\begin{table}[t]
    \centering
    \small
    \setlength{\tabcolsep}{3pt}
    \caption{\textbf{Module ablation.} Contribution of each component.}
    \label{tab:module}
    \resizebox{0.46\textwidth}{!}{
        \begin{tabular}{lc cc cc}
            \toprule
            Variant & Acc & P & R & F1 & WF1 \\
            \midrule
            \textbf{Full model}& 81.57& 79.16 & 79.85 & 79.18 & 81.53  \\
            \midrule

            ~~Only Encoder & 77.98 & 75.22& 75.34& 74.52 & 77.50 \\
            
            ~~ w/o contextual router      &80.49 & 77.90&78.23&	77.63& 80.38  \\
            ~~ w/o generative restoration & 80.49	&77.57 & 78.58 & 77.78 & 80.41  \\
            ~~ w/o precision fusion (uniform) & 80.94 &	78.06 &	79.01 &78.18 & 80.86 \\
            \bottomrule
        \end{tabular}
    }
\end{table}

%% file: tables/diagnostics.tex
\begin{table}[t]
    \centering
    \setlength{\tabcolsep}{4pt}
    \caption{\textbf{Diagnostic-probe study.} Effect of each diagnostic family on the
    weakness estimator.}
    \label{tab:diag}
    \resizebox{0.43\textwidth}{!}{
        \begin{tabular}{lc cc cc}
            \toprule
            Diagnostic set & Acc & P & R & F1 & WF1  \\
            \midrule
            \textbf{Full model} &81.57 &79.16 &79.85 &	79.18 & 81.53  \\
            \midrule
            ~~ $-$ energy (conf)      & 80.22 &	77.72 &	78.71 &77.79 &80.41  \\
            ~~ $-$ epistemic (epi)    & 80.13 & 77.76 & 78.53 &77.63 & 80.15 \\
            ~~ $-$ consensus (rel)    & 80.63 & 77.71 & 77.63 &77.44 &80.63  \\
            ~~ $-$ norm (mag)         &80.33 & 77.73 & 78.29 &	77.62 &80.40  \\
            \midrule
            ~~ conf only              & 81.30 &79.04 &	79.41 &	78.95 & 81.32 \\
            ~~ epi only               & 80.90 & 78.23 &	79.32 &	78.31& 80.83  \\
            ~~ rel only               & 80.76&	78.09&79.48 &	78.35 & 80.73 \\
            ~~ mag only               & 81.17 &	78.67 & 79.29 &	78.62  & 81.16  \\
            \bottomrule
        \end{tabular}
    }
\end{table}

%% file: tables/modality_study.tex
\begin{table}[t]
    \centering
    \small
    \setlength{\tabcolsep}{3pt}
    \caption{\textbf{Modality study.} Effect of which modalities are generatively restored. \checkmark\ marks an enhanced modality.}
    \label{tab:modality}
    \resizebox{0.43\textwidth}{!}{
        \begin{tabular}{lcccccccc}
            \toprule
            Restoration & T & A & V & Acc & P & R & F1 & WF1 \\
            \midrule
            All          & \checkmark & \checkmark & \checkmark  & 81.57 &79.16 &79.85 & 79.18 & 81.53  \\
            Only Text       & \checkmark & & & 80.40 & 77.13 & 78.51 & 77.52 & 80.35 \\  
            Only Audio       & & \checkmark &   &  80.54 & 77.17 & 78.52 & 77.56 & 80.55\\
            Only Video       & & & \checkmark & 80.63 & 78.28 & 78.06	& 77.94 & 80.71 \\
           
             w/o Text     & & \checkmark & \checkmark & 80.40 & 77.85 & 78.14 & 77.78 & 80.34 \\
            w/o Audio    & \checkmark & & \checkmark & 80.99 & 78.18 & 79.22 & 78.42 & 80.89\\
            w/o Video    & \checkmark & \checkmark & & 80.49 & 77.87& 78.33 & 77.84 & 80.42\\
            
            
            \bottomrule
        \end{tabular}
    }
\end{table}

%% file: tables/loss_study.tex
\begin{table}[t]
    \centering
    \setlength{\tabcolsep}{4pt}
    \caption{\textbf{Loss ablation.} Impact of each training objective.}
    \label{tab:loss}
    \resizebox{0.38\textwidth}{!}{
        \begin{tabular}{lccccc}
            \toprule
            Objective & Acc & P & R & F1 & WF1 \\
            \midrule
            \textbf{Full model} &81.57&79.16 &79.85 & 79.18 & 81.53  \\            \midrule
            ~~ w/o $\mathcal{L}_{\text{het}}$   & 79.73 & 77.03 &	78.98 & 77.63 &79.83  \\
            ~~ w/o $\mathcal{L}_{\text{sev}}$   &80.31 & 77.74 & 78.58 & 77.77&80.36 \\ \midrule
            ~~ w/o $\mathcal{L}_{\text{res}}$   & 80.90 &	78.12 & 78.71 & 78.08 &80.79  \\
            ~~ w/o $\mathcal{L}_{\text{gc}}$    & 80.67 &	77.74& 78.66 & 77.92 & 80.65 \\
            ~~ w/o $\mathcal{L}_{\text{kl}}$ & 80.64 &	77.96& 78.96 & 78.12&80.63  \\ \midrule
            ~~ w/o $\mathcal{L}_{\text{lb}}$    & 81.21 &  	78.81& 79.58 & 78.86 & 81.16 \\
             ~~ w/o $\mathcal{L}_{\text{pc}}$    &81.35 & 78.89 & 79.31 & 78.77 &81.25  \\
            \midrule
            ~~ Only $\mathcal{L}_{\text{cls}}$  & 80.72 & 78.10&79.14  & 78.27 & 80.68 \\
            \bottomrule
        \end{tabular}
    }
\end{table}

%% file: 5_conclusion.tex
\section{Conclusion}\label{sec:conclusion}

We presented \textbf{PRIME}, a diagnostic-driven framework for multimodal intent recognition that explicitly estimates modality reliability, restores unreliable representations, and re-estimates reliability before prediction. Instead of treating unreliable modalities as missing or uniformly fusing all modalities, PRIME enables each modality to contribute according to its estimated reliability through a closed-loop \emph{diagnose-restore-reassess} paradigm. Extensive experiments on two multimodal intent recognition benchmarks demonstrate that PRIME consistently improves recognition performance while remaining robust under noisy, missing, conflicting, and text-dominant conditions. The ablation and robustness studies further show that diagnostic reliability estimation, reliability-guided restoration, and precision-weighted fusion each contribute to the overall performance, validating the effectiveness of the proposed framework. More broadly, this work suggests that explicitly reasoning about modality reliability before fusion provides a practical alternative to increasingly complex multimodal interaction models. By combining reliability estimation with targeted restoration, PRIME offers a scalable and interpretable direction for building robust multimodal systems. In future work, we plan to extend PRIME to additional modalities and broader multimodal understanding tasks, including multimodal healthcare and affective computing.